[1994/12/01]% LaTeX date must December 1994 or later
\documentclass[10pt]{article}

\title{Binary Morphisms to Ultimately Periodic Words}
\author{Brendan Lucier\thanks{School of Computer Science, University of Waterloo. Email: \texttt{blucier@alumni.uwaterloo.ca}. Supported in part by NSERC.}}
\date{September 13, 2004}

\usepackage{amsmath,amsthm,fullpage}

\chardef\bslash=`\\ % p. 424, TeXbook
\newtheorem{thm}{Theorem}[section]

\newtheorem{prob}[thm]{Problem}

\newtheorem{prop}[thm]{Proposition}

\theoremstyle{definition}

\theoremstyle{remark}

\newcommand{\eval}[2][\right]{\relax
  \ifx#1\right\relax \left.\fi#2#1\rvert}

\let\abs=\envert

\begin{document}
\maketitle
\markboth{Binary Morphisms to Ultimately Periodic Words}
{Binary Morphisms to Ultimately Periodic Words}
\renewcommand{\sectionmark}[1]{}

\abstract{
This paper classifies morphisms from $\{0,1\}$ that map to ultimately periodic words. In particular, if a morphism $h$ maps an infinite non-ultimately periodic word to an ultimately periodic word then it must be true that $h(0)$ commutes with $h(1)$.
}

\section{Introduction}

In this short note we present a rough solution to an open problem in the study of combinatorics on words, due to Jean-Paul Allouche \cite{Cor}.  We omit a full discussion of the subject matter in this manuscript; there are numerous excellent texts that provide background on the subject of combinatorics on words, and specifically the study of morphisms \cite{Text,Text2}.  The problem of interest is the following:

\begin{prob}
Let $w$ be an infinite word over $\{0,1\}$ that is not ultimately periodic, and let $h$ be a morphism. Suppose $h(w)$ is ultimately periodic. Prove (or disprove) that $h(0)$ commutes with $h(1)$.
\end{prob}

This problem has applications to continued fraction expansions \cite{Roy}. In this paper we shall prove that, indeed,  $h(0)$ must commute with $h(1)$.

A few short notes on notation: given a string $x$, we shall use $x^{\omega}$ to denote the infinite repetition of $x$ (i.e. the word $xxx\cdots$). Also, we shall write $x[i]$ to mean the character of $x$ at index $i$, and $x[i,j]$ to mean the substring of $x$ consisting of those characters at indexes $i$ to $j$, inclusive.  For example, if $x$ is the binary word $0100110$ then $x[2,5] = 1001$.

\section{Main Result}

Before proving the main result, we require a simple proposition regarding injective morphisms.

\begin{prop}\label{prop1}
Suppose $h$ is a morphism from $\{0,1\}$ such that $h(0)$ and $h(1)$ do not commute. Then for all $a,b \in \{0,1\}^*$, $a \neq b \implies h(a) \neq h(b)$.
\end{prop}
\begin{proof}
Assume that $h$ maps to an alphabet that does not contain $0$ or $1$. Denote $x = h(0)$ and $y = h(1)$.
First note that $h(01) \neq h(10)$ implies that $x \neq y$.
Also, if we had $k := \abs{x} = \abs{y}$, then given $h(a)$ we could uniquely determine $a$ simply by matching the letters of $h(a)$ to the letters of $x$ and $y$, $k$ at a time. Being able to uniquely determine $a$ from $h(a)$ would imply that $a \neq b \implies h(a) \neq h(b)$, as required. So we can assume $\abs{x} \neq \abs{y}$. 

Let $n = \max\{\abs{x},\abs{y}\}$. Assume without loss of generality that $\abs{y} > \abs{x}$, so we have $n = \abs{y}$.
We now proceed by induction on n.

If $n = 1$ then we must have $x = \epsilon$. But then we would have $xy = y = yx$, a contradiction.

If $n = 2$ then $\abs{x} = 1$ and $\abs{y} = 2$. But since $x$ and $y$ do not commute, we cannot have $y = xx$. Let $c$ be the first letter that appears in $y$ that is not $x$.
Now, given a string $h(a)$, we can uniquely determine the value of $a$ as follows. Scan the word $h(a)$ from left to right. Every time we find a $c$, we map it plus the preceeding character (if $y = xc$) or the following character (otherwise) to $1$. Once that is done, map all of the remaining $x$ characters to $0$. This process generates the string $a$ in a unique way. We conclude that if $a \neq b$ then we must have $h(a) \neq h(b)$.

This concludes the base cases.

So suppose now that $\max\{\abs{x},\abs{y}\} = n > 2$. Suppose also for contradiction that there exist binary words $a,b$ such that $a \neq b$ but $h(a) = h(b)$. Since $h(a) = h(b)$ and neither $x$ nor $y$ is $\epsilon$ it cannot be the case that either $a$ or $b$ is a prefix of the other. There must therefore be some minimal index $i \leq \min\{\abs{a}, \abs{b}\}$ such that $a[i] \neq b[i]$. But then if we let $z = h(a[1,i-1])$ we have that both $zx$ and $zy$ are prefixes of $h(a)$. We conclude that $x$ is a prefix of $y$. Say $y = xy'$, with $\abs{y'} < \abs{y}$. 

Let $f\colon \{0,1\}^* \to \{0,1\}^*$ be the morphism $f(0) = 0$, $f(1) = 01$. Let $h$ be the morphism on $\{0,1\}^*$ given by $h(0) = x$, $h(1) = y'$. Note that $h = g\circ f$. By our base case, $a \neq b \implies f(a) \neq f(b)$. Also, $\max\{\abs{x}, \abs{y'}\} < \abs{y} = n$, so by induction we must now have 
\begin{equation}
h(a) = g(f(a)) \neq g(f(b)) = h(b)
\end{equation}
as required.
\end{proof}

We are now ready to prove the main result.

\begin{thm}
Suppose $w$ is an infinite word over $\{0,1\}$ that is not ultimately periodic, and let $h$ be a morphism. If $h(w)$ is ultimately periodic then $h(0)$ commutes with $h(1)$.
\end{thm}

\begin{proof}
Suppose for contradiction that $h(0)$ does not commute with $h(1)$. Since $h(w)$ is ultimately periodic, we can write 
\begin{equation}
h(w) = yz^{\omega}
\end{equation}
for finite strings $y$ and $z$.

Note that every prefix of $w$ must map to a prefix of $yz^{\omega}$, so in particular there must be infinitely many prefixes of $w$ that map to a string of the form $yz^*z[1,k]$ for any $k \leq \abs{z}$. But there are only finitely many possible values for $k$. There must therefore be some prefix $z_1$ of $z$ such that infinitely many prefixes of $w$ map to strings of the form $yz^*z_1$ for any $t \geq 0$. 

Now say $z = z_1z_2$. Say that $x, xa_1, xa_1a_2, \dotsc$ are the infinitely many prefixes of $w$ discussed above, where each $a_i$ is chosen to minimize $\abs{a_i}$. Then we have
\begin{equation}
h(a_i) = z_2z^{p_i}z_1 \,\,\forall i \geq 1
\end{equation}
where $p_i \geq 0$ for all $i \geq 1$. 

Suppose first that all $p_i$ are equal. Then since $h(0)$ doesn't commute with $h(1)$, Proposition \ref{prop1} tells us that since all $h(a_i)$ are equal, all $a_i$ must be equal. But then $w = xa_1a_1a_1\dotsm$ is ultimately periodic, a contradiction.

Suppose instead not all $p_i$ are equal, so there is some $i$ such that $p_i \neq p_{i+1}$. Then we have
\begin{equation}
\begin{split}
& h(a_ia_{i+1})\\
= & z_2z^{p_i}z_1z_2z^{p_{i+1}}z_1\\
= & z_2z^{p_{i+1}}z_1z_2z^{p_i}z_1\\
= & h(a_{i+1}a_i)
\end{split}
\end{equation}
So by Proposition \ref{prop1}, $a_ia_{i+1} = a_{i+1}a_i$. 

The second theorem of Lyndon and Schutzenberger now tells us that there exists some $b$ such that $a_i = b^k$, $a_{i+1} = b^l$. Since we assumed $p_i \neq p_{i+1}$ we must have $k \neq l$. If $k < l$ then $\abs{a_i} < \abs{a_{i+1}}$ and $a_{i+1}$ has $a_i$ as a strict prefix. But $a_i$ is of the form $z_2z^*z_1$, so this contradicts the assumed minimality of $\abs{a_{i+1}}$. If $k > l$ then an identical argument contradicts the minimality of $\abs{a_i}$. 

\end{proof}

\end{document}